\documentclass[twocolumn,superscriptaddress,a4paper,nofootinbib,amsmath,amssymb,aps,floatfix]{revtex4-1}
\usepackage{graphicx,color}
\usepackage[normalem]{ulem}
\usepackage{epstopdf}
\pdfoutput=1

\newcommand{\tn}{$T_{\text{N}}$}
\newcommand{\ea}{{\it et al.}}
\newcommand{\mm}[1]{\mbox{$#1$}}
\newcommand{\mnl}{Mn~$L_{2,3}$-edge}

\usepackage{bm}

\begin{document}

\title{Spin-spiral state of a Mn monolayer on
  W(110) studied by soft x-ray absorption spectroscopy at
  variable temperatures}

\author{J.~Honolka}
\email{honolka@fzu.cz}
\affiliation{FZU -- Institute of Physics, Academy of Sciences of the
  Czech Republic, Na~Slovance~2, CZ-182~21 Prague, Czech Republic} 

\author{S.~Krotzky} \affiliation{Max Planck Institute for Solid State
  Research, Heisenbergstrasse 1, D-70569 Stuttgart, Germany}

\author{M.~Menzel} \affiliation{Department of Physics,
  University of Hamburg, Jungiusstr.~11A, D-20355 Hamburg, Germany}

\author{T.~Herden} \affiliation{Max Planck Institute for Solid State
  Research, Heisenbergstrasse 1, D-70569 Stuttgart, Germany}

\author{V.~Sessi} \affiliation{European Synchrotron Radiation Facility
  (ESRF), 71 Avenue des Martyrs, F-38000 Grenoble, France}

\author{H.~Ebert} \affiliation{Department of Chemistry,
  Ludwig-Maximilians-Universit\"{a}t M\"{u}nchen Butenandtstr.~11,
  D-81377 M\"{u}nchen, Germany}

\author{J.~Min\'{a}r} \affiliation{New Technologies Research Centre,
  University of West Bohemia, Univerzitn\'{i} 8, CZ-301~00 Pilsen, Czech Republic}

\author{K.~von~Bergmann} \affiliation{Department of Physics,
  University of Hamburg, Jungiusstr.~11A, D-20355 Hamburg, Germany}

\author{R.~Wiesendanger} \affiliation{Department of Physics,
  University of Hamburg, Jungiusstr.~11A, D-20355 Hamburg, Germany}

\author{O.~\v{S}ipr} \email{sipr@fzu.cz} \affiliation{FZU -- Institute
  of Physics ASCR, Cukrovarnick\'{a}~10, CZ-162~53 Prague, Czech
  Republic}

\date{\today}



\begin{abstract}
The noncollinear magnetic state of epitaxial Mn monolayers on tungsten
(110) crystal surfaces is investigated by means of soft x-ray
absorption spectroscopy, to complement earlier spin-polarized STM
experiments.  X-ray absorption spectra (XAS), x-ray linear dichroism
(XLD) and x-ray magnetic circular dichroism (XMCD) \mnl\ spectra were
measured in the temperature range from 8 to 300~K and compared to
results of fully-relativistic {\em ab initio} calculations.  We show
that antiferromagnetic (AFM) helical and cycloidal spirals give rise
to significantly different \mnl\ XLD signals, enabling thus to
distinguish between them.  It follows from our results that the magnetic
ground state of a Mn monolayer on W(110) is an AFM cycloidal spin
spiral.  Based on temperature-dependent XAS, XLD and field-induced
XMCD spectra we deduce that magnetic properties of the Mn monolayer on
W(110) vary with temperature, but this variation lacks a clear
indication of a phase transition in the investigated temperature range
up to 300~K --- even though a crossover exists around 170~K in the
temperature dependence of XAS branching ratios and in XLD profiles.
\end{abstract}


\maketitle


\section{Introduction \label{sec:introduction}}

Noncollinear magnetic states have attracted a lot of attention lately.
However, experimental investigations of noncollinear magnetism are
difficult: standard spatial-averaging techniques cannot be used
because the average magnetization is zero.  New techniques have to be
sought.  It was demonstrated that scanning tunneling microscopy (STM)
with magnetic or nonmagnetic tips can provide valuable information
about local magnetic order \cite{BHK+02}. In particular, a Mn
monolayer on W(110) [called Mn/W(110) hereafter] has become a
playground for studying noncollinear magnetism. Ordered Mn monolayer
stripes can be prepared on stepped W(110) single crystal substrates by
epitaxial step-flow growth \cite{Bode2002}.  First studies of the
magnetic order of Mn/W(110) by spin-polarized STM experiments
\cite{HBK+00} as well as {\em ab initio} calculations
\cite{HBK+00,Dennler2005} suggested a collinear antiferromagnetic
(AFM) order.  Later, field-dependent spin-polarized STM experiments
with atomic resolution for a larger field of view revealed a periodic
magnetic pattern along the [1$\bar{1}$0] direction of W(110), with a
wavelength $\lambda$ of about 12~nm, indicating an AFM cycloidal or
helical spin-spiral magnetic ground state \cite{BHB+07}.  Theoretical
investigations accounting for the Dzyaloshinskii-Moriya interaction
proposed that the ground state should be a left-handed AFM cycloidal
spin spiral \cite{BHB+07}.  Later, additional field-dependent STM
measurements provided further evidence in favor of this
\cite{Haze2017,Haze2017}.  Recently the Mn/W(110) system was exploited
as a substrate inducing noncollinear magnetism in adatoms
\cite{SYM+16,HH+18,GHH+20}.

The temperature dependence of the magnetic order of Mn/W(110) is
especially interesting.  Estimates of the N\'{e}el temperature
\tn\ based on theoretical and experimental studies differ
substantially: atomistic simulations of the magnetic order with the
input parameters taken from {\em ab initio} calculations yield \tn\ of
about 510~K whereas STM studies exploiting the spin-orbit contrast
suggest \tn\ to be about 240~K \cite{Hasselberg2015}, with additional
dependence of \tn\ on the structural width of Mn monolayer stripes
along the [001] direction \cite{Sessi2009}.  These results could be
reconciled by considering different time scales probed by the
experiment in connection with thermal depinning of the spin
spirals. Neither theory nor STM experiment found any indication for an
intermediate magnetic state between the cycloidal spin spiral (CSS)
state and a paramagnetic state at higher temperatures.

Even though spin-polarized STM is a very powerful method, it may not
provide the full picture.  In particular, if a thermal depinning takes
place for a spin spiral state, it is possible that the spin spiral
moves along its propagation direction, resulting in a phase
shift. When this process is fast compared to the time-scale of the
measurement, i.e., the settling time of the STM tip per pixel, only a
time-averaged signal is obtained \cite{MMW+12,Hasselberg2015}.
Regarding spatial information, typical surface imaging areas of STM
are up to few hundreds of nanometers, which is a limitation when a
large variety of Mn stripe geometries exists and good statistics is
achieved only on larger areas.  Employing a complementary technique
which would provide instantaneous time snapshots for a representative
part of the sample is thus desirable.
  
A suitable technique in this respect is x-ray absorption spectroscopy.
It is chemically specific, so one knows that one gets a view on the
electronic states from an atom of a given type and, at the same time,
the inspected area is macroscopic (typically 0.1~mm$\times$0.3~mm).
Dichroic techniques studying the change of x-ray absorption spectrum
(XAS) upon changing the polarization vectors of the incoming x-rays
and/or the direction of the magnetization proved to be powerful for
studying ferromagnetic as well as antiferromagnetic order.  For
studying complicated noncollinear magnetic order such as the one
formed in Mn/W(110), XAS spectroscopy was used only scarcely
\cite{KCO+04,Grazioli2005,LRB+19}. However, dealing with its
complexity is worthy because x-ray absorption spectroscopy can be
viewed as a complementary technique to STM.
  
So far x-ray magnetic linear dichroism (XMLD) techniques were mostly
applied to collinear AFM bulk systems where there would be no dichroic
signal without magnetic order --- see, e.g.,
Refs.~\cite{NSS+00,ALC+06,WHS+15}.  In such cases the orientation of
x-ray polarization vectors with respect to the crystal affects the
shape and intensity of the XMLD spectra \cite{KO03a,ALC+06}, but the
very reason for the effect is magnetic.  When studying Mn/W(110), one
encounters a different situation: there is also a strong structural
component of the dichroism, stemming from the non-fourfold symmetry of
the bcc (110) crystal surface. The magnetic order is an additional
factor affecting the spectrum.

Our study focuses on measuring temperature-dependent XAS at the Mn
$L_{2,3}$ edges and the respective x-ray linear dichroism (XLD) and
field-induced x-ray magnetic circular dichroism (XMCD) for Mn/W(110)
and on comparing these data to {\em ab initio} calculations.  This
enables us to assess to what extent the XLD technique can be applied
to distinguish between various noncollinear magnetic configurations.
In particular, we find that, for the particular case of Mn/W(110), the
XLD technique is sensitive to the differences between cycloidal and
helical spin spirals.  Another important issue we explore is how the
magnetic order of Mn/W(110) changes if the temperature increases from
8~K up to room temperature.  Magnetic properties of Mn/W(110) vary
with temperature, but there is no clear indication of a phase
transition in the investigated range.  Nevertheless, a crossover
around 170~K is apparent in the temperature dependence of $L_{3}$ to
$L_{2}$ XAS branching ratios and in XLD profiles (but not in XMCD);
the exact nature of this crossover is not clear.


\section{Methods \label{sec:methods} } 


\subsection{Sample preparation \label{sec:sample} }

W(110) crystal preparation and Mn deposition were done {\it in situ}
in the UHV preparation chamber of the ID8 beamline at ESRF.  W(110)
single crystals were mounted on standard Omicron plates and cleaned in
a standard two-step process described elsewhere \cite{BHB+07}: the
crystal was annealed at $T = 1200^{\circ}$C in oxygen atmosphere, and
the surface oxide formed during this process is thereafter removed by
a short flash to $T = 1800^{\circ}$C. This cycle was repeated until STM
and low-energy electron diffraction (LEED) showed well-ordered and
clean surface properties.

\begin{figure}
\includegraphics[width=85mm]{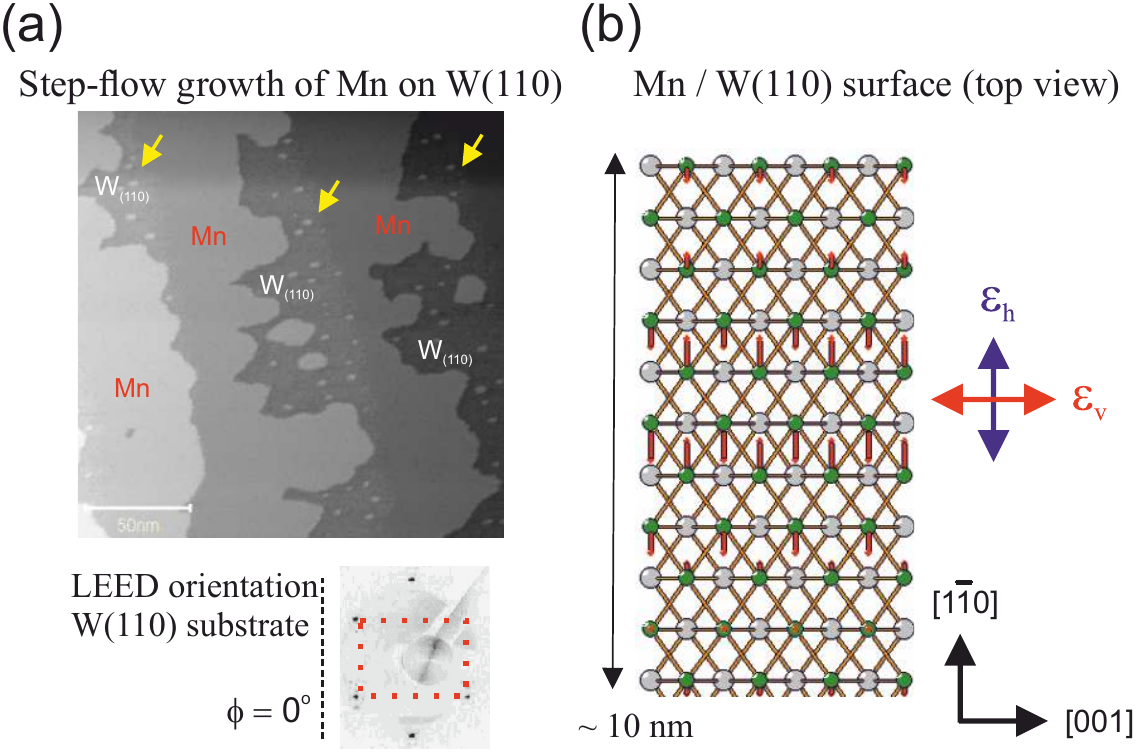} 
\caption{\label{fig:measuring-geometry} (a) STM image of the Mn/W(110)
  surface with a Mn coverage of about 80\% of a ML. Yellow arrows indicate
  W(110) step-edges. The LEED image of the W(110) surface taken before
  Mn growth defines the azimuthal orientation $\phi = 0^{\circ}$ with
  respect to $\epsilon_{h}$ and $\epsilon_{v}$ incoming photon polarization
  vectors. (b)~Schematic depiction of the surface atomic configuration
  and respective orientation of $\epsilon_{h}$ and $\epsilon_{v}$
  photon polarization vectors for $\phi = 0^{\circ}$.  The small
  dark
  circles denote Mn atoms, the large light cirles denote W atoms in
  the sub-surface layer.  Rods at Mn atoms indicate symbolically the
  directions of magnetic moments for an antiferromagnetic CSS.
}
\end{figure}

Mn was evaporated from a crucible with deposition rates of
approximately 0.1~monolayer (ML) per minute (evaporator parameters:
1.9~A, 11.4~mA, 500~V, 300~nA). The deposition rate was calibrated by
STM. During deposition the W crystals were kept at about
$200^{\circ}$C, which leads to a step-flow growth of Mn monolayer
stripes along W step-edges as shown in
Fig.~\ref{fig:measuring-geometry}(a). Coverages and morphology of the
Mn monolayers were highly reproducible in STM when samples were
reprepared from scratch using the same parameters. The homogeneity of
the Mn coverages was verified by respective {\it in situ} XAS data ---
the Mn $L_{2,3}$ white line intensities differed by 4\% at most.
Further STM images of the W(110) substrate and Mn/W(110) sample are
shown in Appendix~\ref{sec:stm}.  After fresh preparation the samples
were immediately transferred {\it in situ} to the XAS chamber at ID8.
  
In order to change the azimuthal orientation (see
Sec.~\ref{sec:measurement}), the W(110) crystal was {\it ex
  situ} rotated on the Omicron sample holder and reentered into the
UHV chambers. The {\it in situ} tungsten cleaning procedures described
in Sec.~\ref{sec:sample} were repeated thereafter, followed by a fresh
Mn monolayer preparation and respective STM characterization.


\subsection{Measurement \label{sec:measurement} }

XAS was measured at the Mn $L_{2,3}$ edges (x-ray energy 630--680~eV)
at the ID8 beamline for temperatures varying from 8~K to 300~K, in the
surface sensitive total electron yield (TEY) mode. XLD was measured in
a polar geometry ($\Theta = 0^{\circ}$) with horizontal
$\epsilon_{h}$ and vertical $\epsilon_{v}$ x-ray polarization vectors
(in the laboratory coordinates) of the Apple~II undulator. A small
magnetic field of $B = 100$~mT was applied during the XLD measurements
to reduce the TEY noise.

The XLD spectra were recorded for two different azimuthal crystal
orientations: (i) The [1$\bar{1}$0] direction of the W crystal was
parallel to the x-ray polarization vector $\epsilon_{h}$ [$\phi =
  0\pm2^{\circ}$ --- as depicted in
  Fig.~\ref{fig:measuring-geometry}(b)], and (ii) the direction
W[1$\bar{1}$0] forms an angle $\phi = 48\pm2^{\circ}$ with
$\epsilon_{h}$. These azimuthal orientations were derived from {\it in
  situ} W(110) LEED patterns [as shown, e.g., at the bottom of
  Fig.~\ref{fig:measuring-geometry}(a)].

The XMCD spectra were measured for two incidence angles, namely,
$\Theta = 0^{\circ}$ (polar incidence) and $\Theta = 70^{\circ}$
(grazing incidence, with the plane of incidence defined by the W[110]
and W[1$\bar{1}$0] directions).  During XMCD measurements an external
magnetic field $B = \pm 5$~T is oriented parallel to the photon beam.
The XMCD signal thus reflects the field-induced average Mn
magnetizations along the respective photon beam directions.

Dichroism signals are defined as
\begin{equation}
  \text{XLD} \: = \:
  \mu(\epsilon_{v}) - \mu(\epsilon_{h})
  \: \equiv \:
  \mu(\epsilon \! \parallel \! \text{W}[001]) -
  \mu(\epsilon \! \parallel \! \text{W}[1\bar{1}0])  \;\; ,
\end{equation}
with $\mu(\epsilon)$ denoting the absorption coefficient for incoming
photon polarization vector $\epsilon$, and
\begin{equation}
\text{XMCD} = \mu(\epsilon_{+}) - \mu(\epsilon_{-}) \;\; .
\end{equation}

The respective average non-dichroic signals are defined as
\begin{equation}
  \text{XAS} \: = \:
  [\mu(\epsilon_{h}) + \mu(\epsilon_{v})]/2 
\end{equation}
for linearly polarized x-rays and
\begin{equation}
  \text{XAS} \: = \:
  [\mu(\epsilon_{+}) +  \mu(\epsilon_{-})]/2 
\end{equation}
for circularly polarized x-rays.  All XAS signals are routinely
normalized to unity at the Mn preedge at 635 eV to compensate
variations of the base TEY signal, e.g., due to different incident
photon beam intensities.

As a quantitative measure of the average moment per Mn atom we will
use the so-called XMCD asymmetry. It is defined as the ratio between
the $L_3$ XMCD peak intensity and the respective $L_3$ XAS peak
intensity,
\begin{equation}
  \text{XMCD asymmetry} \: = \; \text{XMCD}_{L_3} \, / \,
  \text{XAS}_{L_3}
  \;\; ,
\end{equation}
evaluated at the $L_3$ edge (at about 641~eV in our case).

To check the correct photon energy calibration, the signal of a
reference MnO polycrystalline sample was monitored during all
polarization-dependent measurements to verify the absence of XLD in the
reference sample at all times  (see Fig.~\ref{fig:MnO} in
Appendix~\ref{sec:xldcheck}).


\subsection{Calculations \label{sec:calculations} }
        
Accompanying calculations were done within the {\em ab initio}
framework of spin-density functional theory (SDTF), relying on the
generalized gradient approximation using the Perdew, Burke and
Ernzerhof (PBE) functional \cite{PBE96}. The electronic structure
embodied in the underlying effective single-particle Dirac Hamiltonian
was calculated in a fully relativistic mode using the spin-polarized
Green's function multiple-scattering (KKR-GF) formalism \cite{EKM11}
as implemented in the {\sc sprkkr} code \cite{sprkkr-code}.

The Mn/W(110) system was modeled by a slab consisting of six W layers
and a Mn layer, embedded in vacuum (represented by three layers of
empty sites above the Mn layer).  The lattice constant of bcc W was
set to $a_{0} = 3.165$~\AA, and the vertical distance between the Mn
monolayer and the nearest W layer to 2.12~\AA, in agreement with
earlier {\em ab initio} calculations
\cite{Bode2002,Dennler2005}. First, the electronic structure of the
two-dimensional slab was calculated by means of the tight-binding KKR
technique \cite{ZDU+95}, to obtain self-consistent potentials.
Afterwards, the spectra were evaluated by the real space calculations,
using a cluster of 10.40~\AA\ radius containing 181~atoms (or
315~atoms if counting empty vacuum sites as well). We checked that
increasing the cluster further does not lead to significant changes in
the results.  Raw theoretical spectra were broadened by Lorentzians
with full widths at half maxima 0.70~eV ($L_{2}$ edge) and 0.35~eV
($L_{3}$ edge), to account for the finite core hole lifetime.

Spectra for spin spirals were evaluated as averages of spectra for two
AFM configurations, with perpendicular directions of the
magnetization.  In particular, spectra for AFM CSS are evaluated by
averaging spectra for AFM configurations with $\bm{M}\!  \parallel
\! \text{W}[1\bar{1}0]$ and with $\bm{M}\! \parallel \!
\text{W}[110]$ whereas spectra for AFM helical spin spiral are
evaluated by averaging spectra for AFM configurations with
$\bm{M}\! \parallel \! \text{W}[001]$ and with $\bm{M}\!
\parallel \! \text{W}[1\bar{1}0]$.  One can view this approach as
taking the long-wave-length limit.  We checked by explicit
calculations that this approach is justified in our case (see
Appendix~\ref{sec:spirals}).  Note that a similar representation of
CSS with a collinear AFM configuration was employed recently for {\em
  ab initio} calculations of the magnetic exchange force between an
STM tip and Mn atoms on W(110) \cite{HHH+20}.  Dealing with the
long-wave-length limit means that we cannot in principle distinguish
between the rotational senses of the cycloidal spirals (right- or
left-handed).

The potential was without any shape approximation, i.e., a full
potential mode was employed, relying on the representation of the
atomic cells by means of the shape functions \cite{Zel87a,HZE+98}. The
core hole was included within the final state approximation, i.e., the hole
is relaxed and screened.  Technically, this was achieved within the
single-site impurity approach (see Ref.~\cite{SKJ+19} for more details).
Accounting for the full-potential and the core-hole does not lead to
significant changes in XAS, but it is important for the dichroic
spectra.  The influence of the core hole is illustrated in
Appendix~\ref{sec:corehole}.

The angular momentum cutoff used for calculating the spectra was
\mm{\ell_{\mathrm{max}}} = 3.  A lower cutoff of
\mm{\ell_{\mathrm{max}}} = 2 would be sufficient for XAS and XMCD, but
to get reliable results for XLD, \mm{\ell_{\mathrm{max}}} = 3 was
needed.  The need for \mm{\ell_{\mathrm{max}}}\ larger than what is
common for such systems is probably associated with the fact that the
\mnl\ XLD signal is quite small and, therefore, a high accuracy is
needed.  If an even larger cutoff of \mm{\ell_{\mathrm{max}}} = 4 is
used, no significant changes with respect to the
\mm{\ell_{\mathrm{max}}} = 3 case occur.

When interpreting experimental spectra recorded at elevated
temperatures one has to consider a situation when there are local
magnetic moments at Mn atoms but without any order --- we call such a
system paramagnetic.  We model this by a disordered local moment (DLM)
state: a Mn site is occupied with the same probability by an atom with
spin up and by an atom with spin down \cite{SGP+84}. Technically, this
is achieved via the coherent potential approximation (CPA), which can
be conveniently implemented within the KKR-GF formalism.  To simulate
the disorder also concerning the {\em directions} of the magnetic
moments, additional averaging of spectra calculated for the
magnetization oriented along three perpendicular directions ($\bm{M}
\! \parallel \!  \text{W}[001]$, $\bm{M}\!  \parallel \!
\text{W}[1\bar{1}0]$, $\bm{M}\! \parallel \! \text{W}[110]$) is
performed.  Finally, we dealt also with nonmagnetic Mn; in such a case
the potentials for the spin-up and spin-down electrons are identical
and the local magnetic moments are zero.


\section{Results \label{sec:results}}


\subsection{Magnetic ground state: cycloidal versus
  helical spirals}

\label{sec:lowtemp}

\begin{figure}
\includegraphics[bb=10 15 258 286,width=83mm]{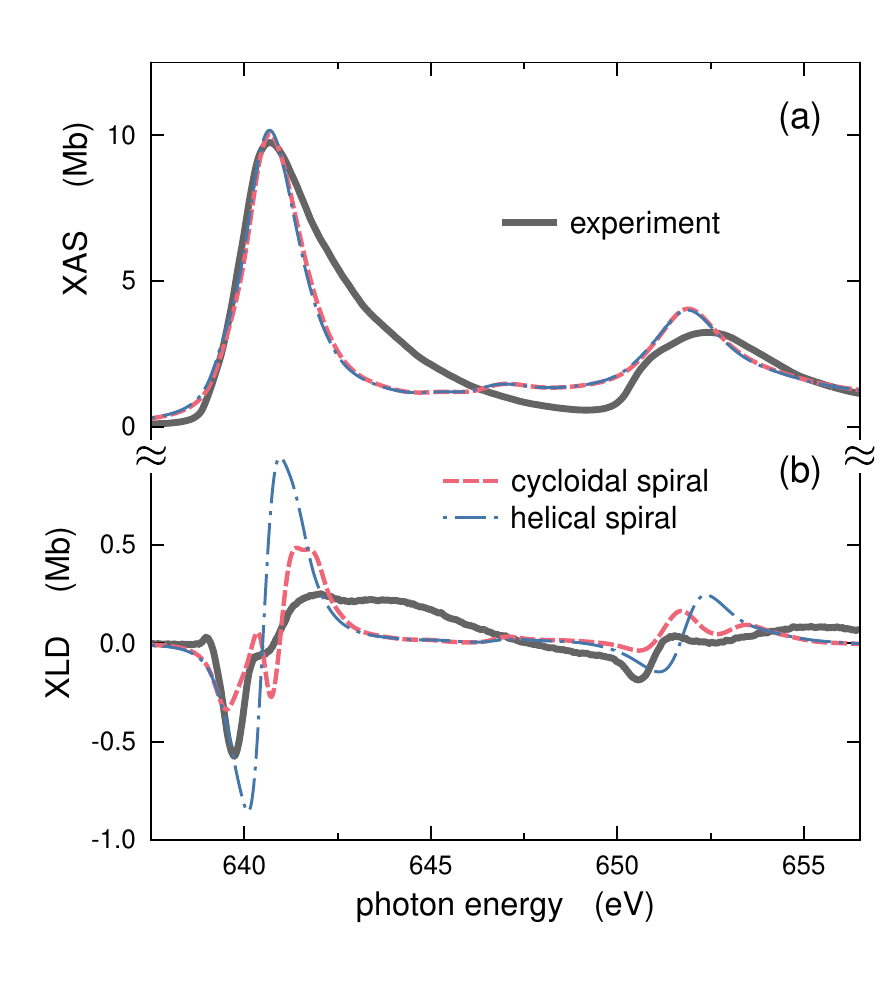}
\caption{\label{fig:xasxld} (a) Experimental XAS and (b) XLD
   spectrum recorded at 8~K compared to spectra
  calculated for AFM cycloidal spin spiral and AFM helical spin spiral
  states. }
\end{figure}

Experimental \mnl\ XAS and XLD spectra of Mn/W(110) recorded at $T =
8$~K in normal incidence geometry ($\Theta = 0^{\circ}$) for azimuthal
orientation of the tungsten crystal $\phi = 0^{\circ}$ are shown in
Fig.~\ref{fig:xasxld} (solid grey lines).  The XAS spectral shape
looks broad, typical for Mn atoms in a metallic state
\cite{SZS+15}. At the $L_2$ edge a faint shoulder is visible at around
651~eV.  The fact that the spectrum resembles spectra of Mn in a
metallic state suggests that atomic-like multiplet effects will be
less significant.  This is even more probable for XLD where the
extended aspects of electronic states are important.  Using an {\em ab
  initio} theoretical scheme, which accounts for the delocalized nature
of electron states, thus seems plausible.

We put the experimental XLD to test by verifying that it conforms the
fundamental $C_{2v}$ symmetry of the system.  To achieve this, we
compare the XLD signals measured at two different azimuthal crystal
orientations, $\phi = 0^{\circ}$ and $\phi = 48^{\circ}$.  The signal
recorded at $\phi = 48^{\circ}$ scales very accurately with the signal
recorded at $\phi = 0^{\circ}$, with the scaling factor $\cos(2\phi) =
-0.105$ (Fig.~\ref{fig:aziscale} in
Appendix~\ref{sec:xldcheck}).  The measured signal thus indeed
corresponds to the linear dichroism effect.

The theoretical spectra shown in Fig.~\ref{fig:xasxld} were calculated
assuming a CSS and a helical spin spiral configuration. The spectra
were aligned in energy so that the theoretical $L_{3}$-edge white line
maximum coincides with experiment.  On the other hand, the vertical
scaling of the spectra was done by scaling the experimental XAS
spectrum so that it matches the theoretical spectrum at the
high-energy tail (because the theory provides absolute units for the
XAS cross-section).  XLD spectra were scaled by the same factor as XAS
spectra.  

Theoretical XAS spectra (practically identical for both magnetic
configurations) reproduce the gross features of the experiment but
fail to reproduce the asymmetric shape of the white lines.  This seems
to be a general feature of {\em ab initio} calculations of
$L_{2,3}$-edge spectra of transition metals and probably is associated
with deficiencies of the final state approximation in describing the
core hole effect in metals \cite{SMS+11}.  The shoulder at 651~eV
observed in experiment probably stems from multiplet effects: it does
not appear in \mnl\ XAS of elemental Mn, but it is present for
Mn-containing semiconductors and insulators \cite{KKL+08,SZS+15} and
also for Mn monolayers on Ag(001) \cite{Schieffer1999} and on Fe(001)
\cite{Grazioli2005}.  If the coverage is increased, the shoulder at
651~eV gradually disappears \cite{Schieffer1999}.  So we can interpret
these features in Fig.~\ref{fig:xasxld} as indications that the
electron states of Mn have a partially local atomic-like character in
our system and that some aspects of theirs cannot be fully accounted
for within common implementations of the SDFT.  The features are
nevertheless small, meaning that the metallic character prevails.

There is no clear experimental counterpart to the small peak appearing
in the calculated XAS spectrum at 647~eV [Fig.~\ref{fig:xasxld}(a)].  This
resembles the situation for transition metals Fe and Co
\cite{SE05,SMS+11}.  As discussed in \cite{SE05}, it is probably
related to a van Hove singularity.

Concerning the XLD, one can see that cycloidal and helical spin
spirals give rise to significantly different signals
[Fig.~\ref{fig:xasxld}(b)]. The helical spin spiral generates a simple
derivative-like (plus/minus) structure both at the $L_3$- and
$L_2$-edge, whereas the CSS shows a more complex XLD lineshape.
Theoretical data for a CSS reproduce the main features of the
experimental data.  Nevertheless, differences in details remain,
especially further away from the edge.  This may be partially linked
to the presence of minor multiplet effects in the spectra.  Another
factor to consider is the possible presence of a second Mn layer in
minor areas of the sample: although the Mn coverage is below one
monolayer and no double-layer Mn islands are observed in the STM image
of Fig.~\ref{fig:measuring-geometry}(a) 1(a), we cannot exclude seeds
of a second Mn layer in other parts of the probed surface
area. However, we checked a reference sample with coverage as low as
30~\% and the spectral 
shape of the XLD is very similar, suggesting that we really deal with
essentially a monolayer system.

The theoretical XLD spectrum for a helical spiral fits the experiment
much worse than the theoretical XLD spectrum for a cycloidal spiral ---
in particular as concerns the characteristic wiggle at 640.5~eV.  So
we conclude that soft x-ray absorption spectroscopy confirms that the
magnetic ground state of Mn/W(110) is an AFM cycloidal spin spiral.


\subsection{Temperature-dependent XLD}

\label{sec:tempxld}

\begin{figure*}
\includegraphics[width=175mm]{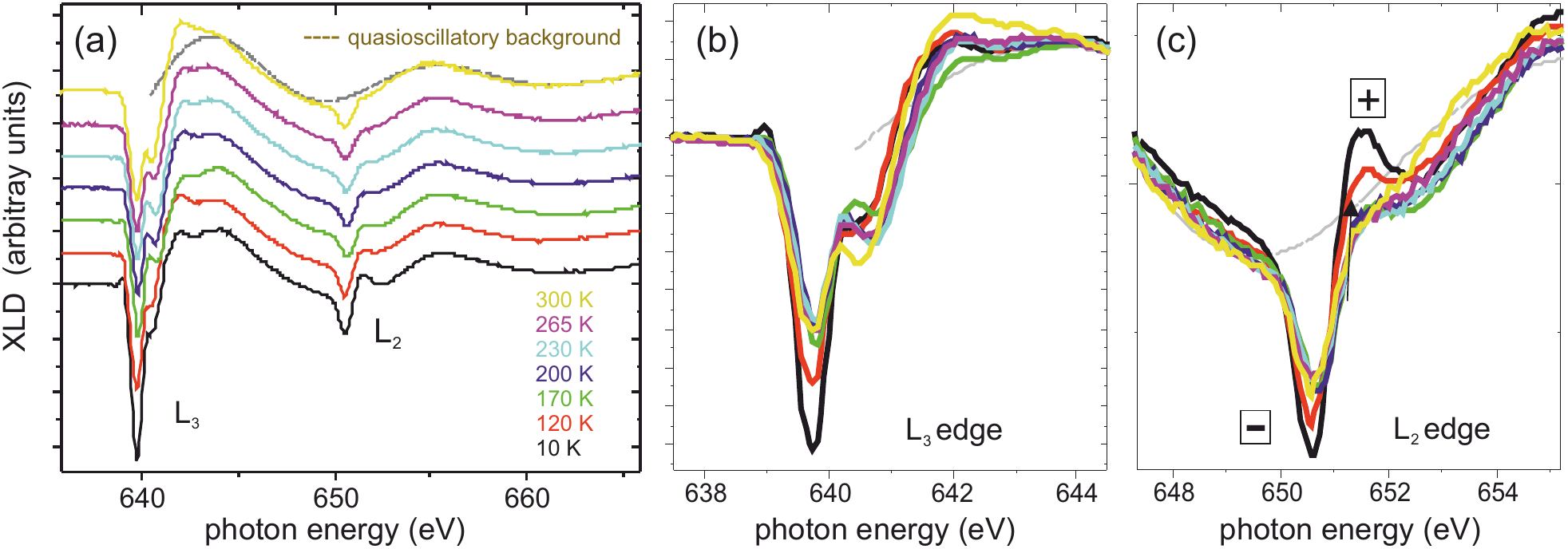}
\caption{\label{fig:T-dependent-xmld} (a) Temperature dependence of
  the XLD signal. The data are shown with a vertical offset for better
  visibility. The dashed curve ontop of the $T = 300$~K curve represents
  a fitted temperature-independent background.  (b)~Detailed view of
  temperature-dependent XLD at the $L_3$ edge, without the vertical
  offset.    (c)~As panel (b) but for the $L_2$ edge.  The minus/plus
  signs indicate features where the temperature dependence of the
  XLD is most pronounced.  }
\end{figure*}

Figure~\ref{fig:T-dependent-xmld} summarizes the temperature
dependence of XLD signals from $T = 8$ to 300~K for the $\phi =
0^{\circ}$ azimuthal orientation of the W crystal.  An overall view is
presented in Fig.~\ref{fig:T-dependent-xmld}(a) (with an {\em ad hoc}
vertical offset between the curves), detailed views on the XLD at the
$L_3$ and $L_2$ edges are shown in Fig.~\ref{fig:T-dependent-xmld}(b)
and Fig.~\ref{fig:T-dependent-xmld}(c) (without offsets).  The data
can be decomposed into a temperature independent broad
quasioscillatory background [the dashed curve in
  Fig.~\ref{fig:T-dependent-xmld}(a)] and temperature dependent
features at the $L_3$ and $L_2$ edges.

The XLD profile changes with temperature.  A closer inspection of the
respective data in Fig.~\ref{fig:T-dependent-xmld} reveals that most
of the change occurs between 8 and 170~K.  Minor changes occur
between 170 and 300~K, especially at the $L_{3}$~edge.  As a whole,
the changes are gradual, without any abrupt jumps which would signal
phase transitions.

\begin{figure}
\includegraphics[bb=10 15 258 320,width=83mm]{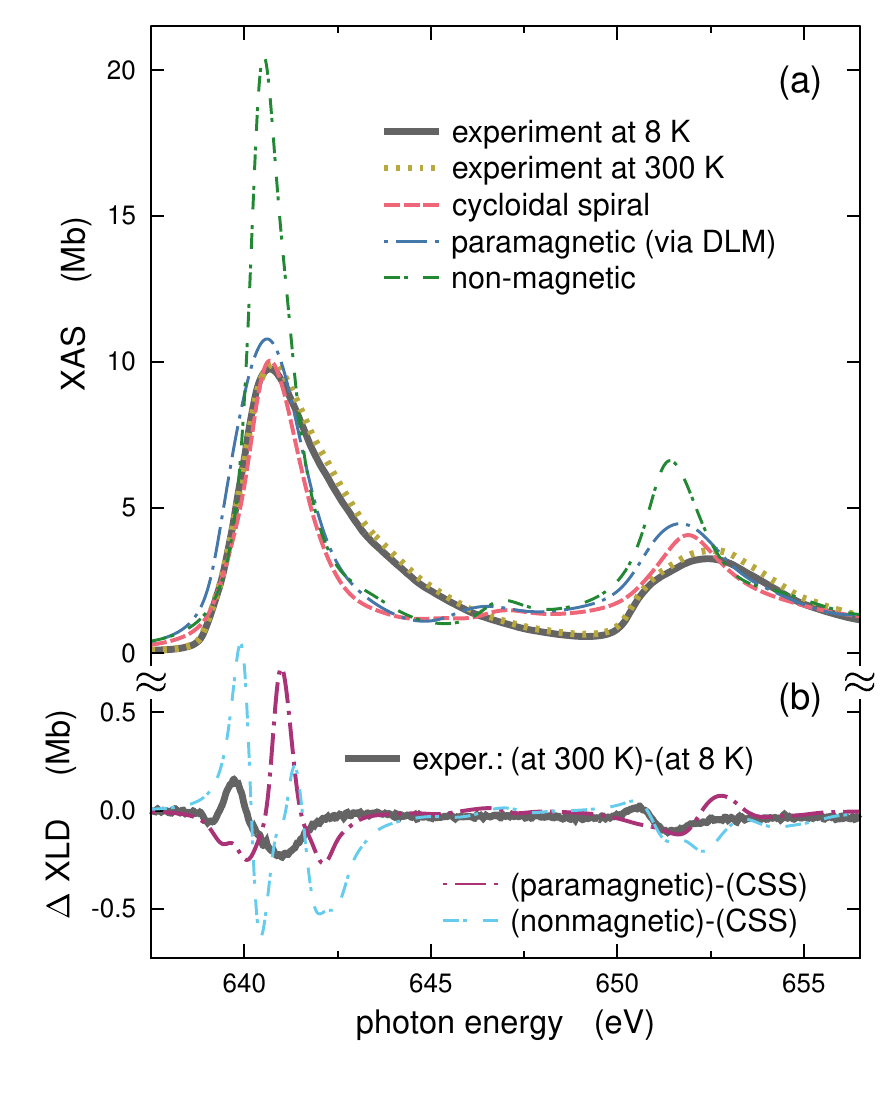}
\caption{\label{fig:dlm} (a) Experimental XAS recorded at 8 and 
  300~K together with theoretical XAS calculated for a CSS state, for
  a paramagnetic state, and for a nonmagnetic state.  (b) Difference
  of experimental XLD signals recorded at 300 and 8~K compared to
  the differences between respective theoretical XLD signals obtained
  by assuming a paramagnetic or non-magnetic state at 300~K 
  and a CSS state at 8~K.}
\end{figure}

To learn how the decay of the magnetic order could possibly influence
the spectra, we calculated \mnl\ XAS and XLD assuming that (i) Mn is
nonmagnetic
and (ii) that Mn is paramagnetic (by averaging DLM spectra for three
perpendicular directions of the magnetization, see
Sec.~\ref{sec:calculations}).  The XAS spectra are shown in
Fig.~\ref{fig:dlm}(a), together with theoretical spectrum for a CSS
and with experimental spectra recorded for $T = 8$ and 300~K.  The
experimental spectra practically do not change if the temperature is
increased.  Likewise, we see only small changes in the theoretical
spectra when going from a CSS to the paramagnetic case.  However, the
\mnl\ XAS for nonmagnetic Mn/W(110) differs quite a lot regarding the
intensity of the $L_{3}$ and $L_{2}$ white lines (even though the
spectra are similar in the high-energy tail region, for $E \gtrsim
655$~eV).  This is in accordance with the changes in the density of
states (DOS): there is a big increase of the DOS just above $E_{F}$ if
the system is made nonmagnetic (see Fig.~\ref{fig:dos} in
Appendix~\ref{sec:dos}).  The fact that there is no big difference
between experimental XAS for $T = 8$ and 300~K can thus be seen as a
proof that there are local magnetic moments present at Mn atoms up to
room temperature.  To a certain degree this is reminiscent of the
situation for Fe and Co, where the local magnetic moments survive up
to high temperature while the magnetic order has already decayed
\cite{RSS+02,PSW+10}.

To isolate the influence of the temperature on XLD, the difference
between XLD signals measured for $T = 8$ and 300~K,
\begin{displaymath}
  \text{XLD}(300\: \text{K}) \; - \; \text{XLD}(8\; \text{K})
  \;\; ,
\end{displaymath}
is presented in Fig.~\ref{fig:dlm}(b).  This
difference is smaller than but comparable to the intensity of the XLD
signals themselves [cf.\ Fig.~\ref{fig:xasxld}(b)]. 
  This suggests that changes indeed happen in the system if the
  temperature increases.  At the same time, the change of the
  experimental XLD between $T = 8$ and 300~K is quite
  different from the change of the theoretical XLD when going from a
  CSS to a paramagnetic state or from a CSS to a nonmagnetic state
  (see the respective differences in  Fig.~\ref{fig:dlm}).
  We can thus infer that at room temperature the system is neither
  nonmagnetic nor paramagnetic.  The exact nature of the
  temperature-induced changes in the magnetic state of Mn/W(110)
  remains unknown.


\subsection{Temperature-dependent field-induced XMCD}

\label{sec:xmcd}

\begin{figure}
\includegraphics[width=85mm]{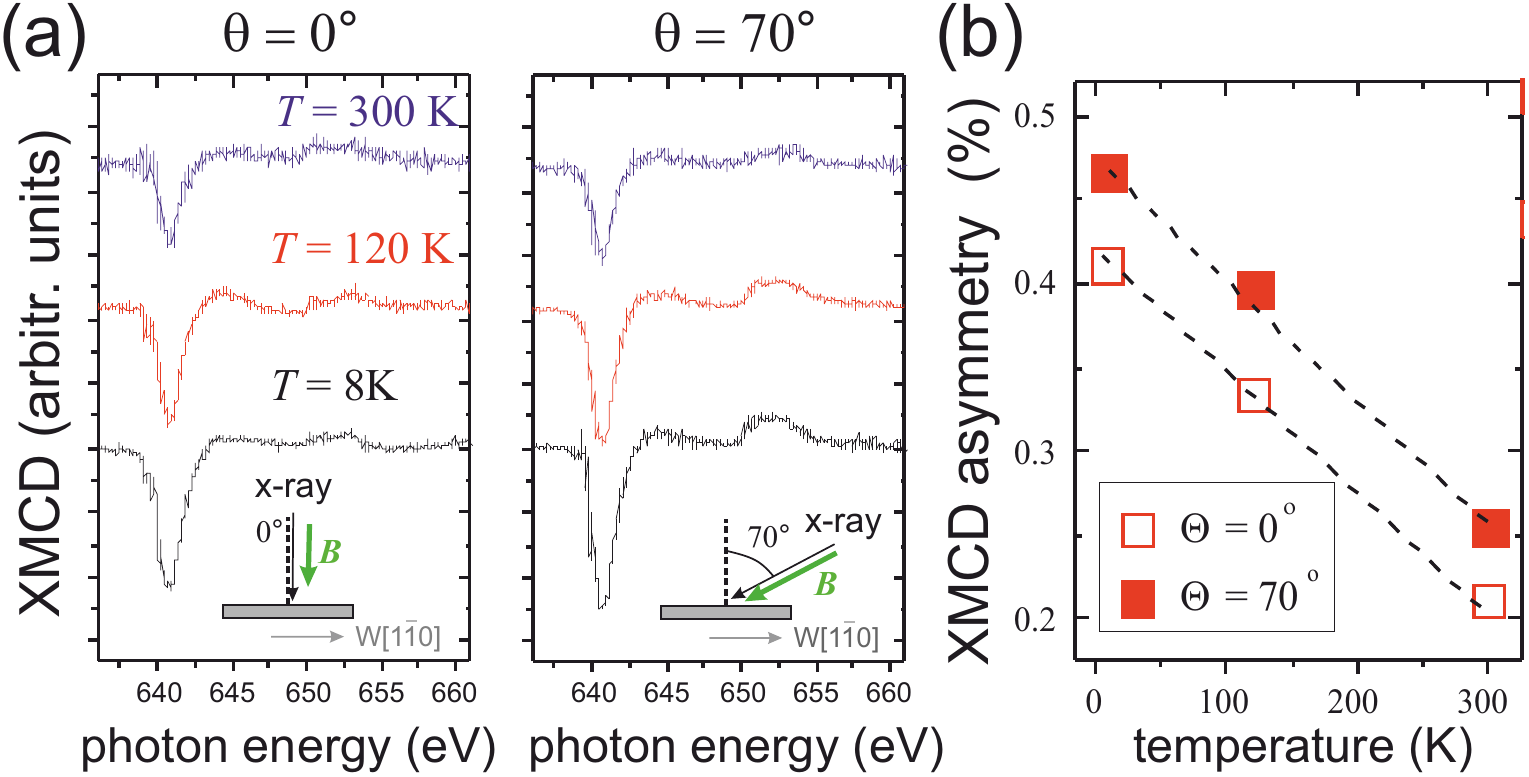}
\caption{\label{fig:xmcd} (a) Experimental \mnl\ XMCD spectra of
  Mn/W(110) for $T = 8$, 200, and 300~K, for $\Theta = 0^{\circ}$
  (normal incidence) and $\Theta = 70^{\circ}$ (grazing incidence).
  An external magnetic field of $B = 5$~T has been applied, as
  indicated in the insets.  (b)~The
  temperature dependence of corresponding XMCD asymmetries, both
  for the normal and for the grazing incidence. }
\end{figure}

The XMCD spectra were measured for polar ($\Theta = 0^{\circ}$, photon
beam parallel to W[110]) and grazing ($\Theta = 70^{\circ}$, photon
beam inclined towards the W[1$\bar{1}$0] direction) incidence. An
external magnetic field of $B = 5$T was applied along the x-rays
incidence direction. The azimuthal orientation of the tungsten crystal
was kept the same as during the XLD measurements, that is $\phi =
0^{\circ}$. The measurements were done for three temperatures ($T =
8$, 200, and 300~K).

The results are summarized in Fig.~\ref{fig:xmcd}:
Fig.~\ref{fig:xmcd}(a) shows the temperature dependent XMCD spectra,
Fig.~\ref{fig:xmcd}(b) presents the respective XMCD asymmetries
derived from the peak heights. The XMCD signal is very small: the XMCD
asymmetry at the $L_{3}$ peak is at most 0.5~\% (at $\Theta =
70^{\circ}$ and $T = 8$~K). For comparison, Mn in a ferromagnetic
state with $\bm{M}$ fully aligned along the x-ray direction would
exhibit an XMCD asymmetry of about 30~\% \cite{HKM+02}. The effect we
observe is a field-induced (non-remanent) XMCD \cite{ME04b}; we
checked that it disappears when the external field is decreased to
zero.   This is in contrast to common XMCD experiments on
  ferromagnets, where the external magnetic field is used just to
  orient the magnetic domains and where, therefore, a nonzero XMCD
  signal remains after the external field has been withdrawn.

\begin{figure}
\includegraphics[bb=10 15 258 263,width=83mm]{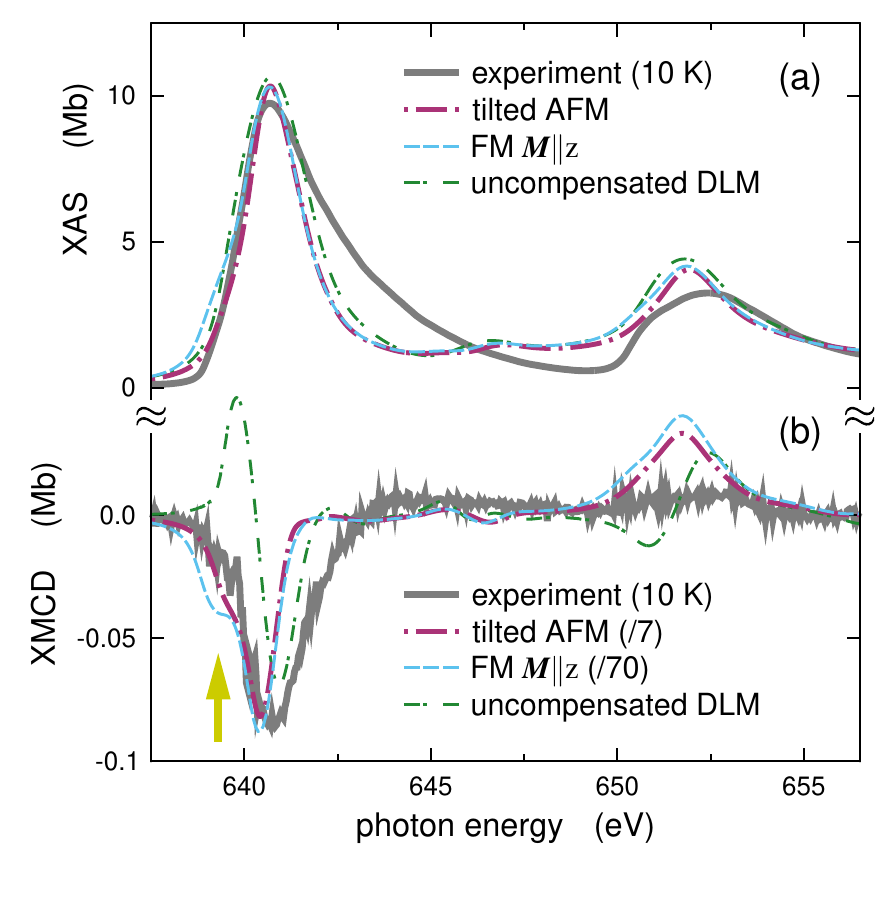}
\caption{\label{fig:tilted} (a) Experimental XAS and (b) field-induced
  XMCD recorded at 8~K compared to theoretical spectra calculated for
  system (i) in an uncompensated DLM state, (ii) in a ferromagnetic
  state, and (iii) in an AFM state with moments nearly in-plane but
  tilted by $5^{\circ}$ in the out-of-plane direction.  Note that the
  XMCD signal for the tilted AFM state was divided by~7 and the XMCD
  signal for the ferromagnetic state was divided by~70.  }
\end{figure}

The XMCD signal is sensitive to the projection of the magnetic moment of
the photoabsorbing atom along the direction of the incoming photons.
Quantitatively, the XMCD intensity $I_{\text{XMCD}}$ scales as
\begin{equation}
I_{\text{XMCD}} \sim \cos\alpha
\label{eq:alpha}
\end{equation}
where $\alpha$ is the angle between the direction of the incoming
x-rays and the magnetic moment \cite{SS06}.  To get a more specific
idea how the measured signal may arise, we calculated the \mnl\ XMCD
of Mn/W(110) for a normal incidence photon beam, considering different
magnetic configurations to model the effect of the external magnetic
field on the arrangement of the magnetic moments.  First, we assume
that the system is nearly antiferromagnetic, with magnetic moments at
Mn atoms oriented in-plane along the W[1$\bar{1}$0] direction but
additionally tilted out-of-plane by 5$^{\circ}$ (in the W[110]
direction).  The second model is a simple ferromagnetic state, with a
magnetization oriented along W[110] (i.e., out-of-plane).  Finally, we
employed an uncompensated DLM model: the magnetic moments are parallel
to W[110], but unlike in the paramagnetic case (dealt with in
Sec.~\ref{sec:tempxld}), there is a slight preference in the spin
orientation (51~\% {\rm vers.}\ 49~\%).  There is no magnetic order in
the third model; it could be viewed as a slightly biased set of
disordered local moments.

Calculated XAS and XMCD spectra for the theoretical models are shown
in Fig.~\ref{fig:tilted}, together with experiment.  Note that the
theoretical XMCD signal was scaled down by a factor of 7 for the
tilted AFM model and by a factor of 70 for the ferromagnetic
configuration. From comparison of the theoretical XMCD with the
experiment it follows that the observed field-induced XMCD is
compatible with the model where the antiferromagnetically ordered
magnetic moments are tilted a bit in the direction of the external
magnetic field.  Note in particular that the faint shoulder on the
low-energy side of the $L_{3}$-edge peak (at 639~eV --- see the arrow
in Fig.~\ref{fig:tilted}) is present both in experiment and in theory
(see the arrow in Fig.~\ref{fig:tilted}).

Interestingly, the ferromagnetic model yields a XMCD signal very similar
to the signal for the tilted AFM model (just ten times more
intensive).  On the other hand, the uncompensated DLM model simulating
a magnetically disordered system with a slight prevalence of magnetic
moments along the external field is incompatible with the
experiment.  As the measured XMCD spectral shape does not change
  significantly between 8 and 300~K [despite the slight changes in
  intensity --- see Fig.~\ref{fig:xmcd}(a)], the incompatibility of the
  uncompensated DLM model gives us strong indication that there is no
  magnetic phase transition to a paramagnetic state within the
  temperature range we explored.

Based on Fig.~\ref{fig:tilted} it seems that the observed XMCD signal
arises from tilting of Mn magnetic moments in the out-of-plane W[110]
direction, induced by the external magnetic field. We can make yet
another quantitative reckoning. By assuming for simplicity that all Mn
moments are oriented in the same direction, we can estimate the angle
$\alpha$ between this direction and the surface normal by means of
Eq.~(\ref{eq:alpha}).  Considering the intensity of the calculated Mn
$L_{3}$-edge XMCD peak for the ferromagnetic model, we get the correct
scaling for $\alpha = 89.2^{\circ}$.  The AFM model with moments
tilted from the in-plane direction leads to $\alpha = 89.3^{\circ}$.
Obviously, both models are very simple --- they disregard the spin
spiral nature of the magnetic ground state.  Nevertheless, the good
agreement between both estimates of $\alpha$ suggests that the idea
that the XMCD spectra are generated through tilting of Mn magnetic
moments by the external field is plausible.

This interpretation is consistent with the dependence of the XMCD
asymmetry on temperature shown in Fig.~\ref{fig:xmcd}(b).  The
decrease of the XMCD asymmetry with increasing temperature results, in
this view, from increased thermal disorder and hence larger resistance
of the spins to be aligned by an external field.  The linear decrease
of the XMCD intensity with temperature presents another argument in
favor of the absence of a magnetic phase transition towards a
paramagnetic state within the experimental range of temperatures: For
such a phase transition one would expect a clear discontinuity in the
susceptibility.

Figure~\ref{fig:xmcd}(b) moreover reveals that for all temperatures
the induced XMCD effect is 20\% smaller at polar incidence ($\Theta =
0^{\circ}$, external magnetic field is out-of-plane) compared to that
of grazing incidence ($\Theta = 70^{\circ}$, external magnetic field
is in-plane). This anisotropy is consistent with an in-plane easy axis
direction along W[1$\bar{1}$0] previously reported experimentally for
Mn monolayer on W(110) and predicted by theory \cite{Bode2002}.
Namely, one can assume that field-induced alignment of magnetic
moments along the hard axis will be less efficient compared to
alignment along the easy axis, resulting thus in a smaller average
projected magnetic moment along the out-of-plane direction.


\subsection{Temperature-dependent XAS branching ratio  \label{sec:BR}}

\begin{figure}
\includegraphics[width=60mm]{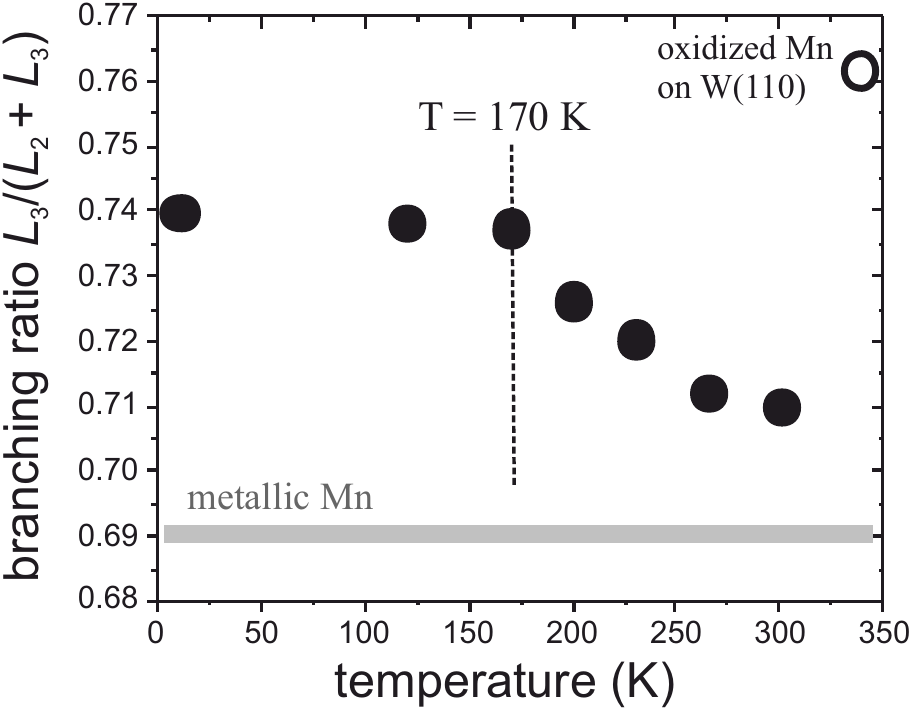}
\caption{\label{fig:branching-ratio} Experimental branching ratios
  $I_{L_{3}}/(I_{L_{2}}+I_{L_{3}})$ for \mnl\ XAS spectra at 
  different temperatures. }
\end{figure}

The ratio of $L_{3}$ and $L_{2}$ intensities in nondichroic XAS
spectra of transition metals contains information on the $d$-shell
spin state.  It can be quantified by the branching ratio (BR) defined
as
\begin{equation}
  \frac{I_{L_{3}}}{I_{L_{2}}+I_{L_{3}}}
  \label{eq:br} \;\; ,
\end{equation}
where $I_{L_{3}}$ and $I_{L_{2}}$ are integrated intensities of the
XAS white lines at the respective edges.  We quantified these
intensities by subtracting a standard two-step-function background
from the measured XAS and integrated the peaks in the energy ranges
637.5--649.5~eV and 649.5--670.0 eV to get $I_{L_{3}}$ and
$I_{L_{2}}$, respectively.

BRs are shown in Fig.~\ref{fig:branching-ratio} for different
temperatures; they were evaluated for \mnl\ spectra recorded for the
$\phi = 0^{\circ}$ azimuthal orientation of the W crystal.  For low
temperatures the BR is constant (BR = 0.74; $I_{L_{3}}/I_{L_{2}}$ =
2.85).  If the temperature increases above $T = 170$~K, the values
drop steadily and at room temperature BR reaches 0.71, which is closer
to that of metallic Mn (BR = 0.69 \cite{Schieffer1999}) and to the
statistical value of 2/3.  For comparison --- the corresponding values
for 14~ML of Mn on fcc Co(001) are BR = 0.72 and $I_{L_{3}}/I_{L_{2}}$
= 2.55 \cite{BT+94}.

According to Thole et al.\ \cite{Thole88} the BR of 3$d$ transition
metals reflects the electronic $d$-shell configuration, which
determines the local moment per atom.  An abrupt change in the
$d$-shell configuration clearly occurs when exposing our Mn monolayer
samples to oxygen: oxygen contamination leads to a characteristic
multiplet structure in the XAS spectral shape corresponding to a
Mn$^{2+}$ state (similar to that of MnO shown in Fig.~\ref{fig:MnO}).
In the oxidized state the BR jumps to higher values of about 0.76 (see
data point at $T = 338$~K in Fig.~\ref{fig:branching-ratio}), which is
consistent with BR's reported for MnO with a $d^{5}$ occupancy
\cite{JLS+09}.
    
The observed small decay of the $I_{L_{3}}/I_{L_{3}+L_{2}}$ branching
ratio between 170 and 300~K (disregarding the oxidized case) suggests
a subtle but continuous change in the Mn $d$-shell spin configuration
with temperature. Interestingly, this happens in the range where the
temperature-induced changes in the XLD peaks are small in comparison
with the changes for $T \lesssim 170$~K (see
Fig.~\ref{fig:T-dependent-xmld}). One should bear in mind that the BR
does not reflect magnetic order but only the magnitude of the local
moments per atom.


\section{Discussion \label{sec:discuss}}

Mn/W(110) is a system where the XLD would arise even without any
magnetism, just because of the non-fourfold symmetry of the bcc (110)
surface.  However, the magnetic state has a profound influence on the
shape of the XLD spectra (see Fig.~\ref{fig:xasxld}).  So in common
terminology one could speak about natural dichroism in the presence of
magnetization.  It is not possible to disentangle the structural and
magnetic contributions to XLD from each other.  This is because
magnetization has a strong influence on the electronic structure, as
demonstrated by comparing the DOS for magnetic and nonmagnetic
Mn/W(110) (Fig.~\ref{fig:dos} in Appendix~\ref{sec:dos}).  Forcing Mn
to be nonmagnetic would change the spectra substantially, similarly as
if Mn was replaced by a different chemical element.

The XLD spectrum depends on the direction of the magnetic moments.
This is crucial for distinguishing between helical and cycloidal spin
spirals. The direction of magnetic moments is connected to the
Mn/W(110) system via spin-orbit coupling (SOC).  The importance of SOC
is illustrated in Appendix~\ref{sec:soc}, where it is shown how XLD
changes if the direction of antiferromagnetically coupled magnetic
moments at Mn atoms varies.  Spin-orbit interaction thus always has to
be considered when analyzing linear dichroism for magnetic systems,
even in cases where an essential part of the dichroism originates from
the structure.

The dependence of magnetism of Mn/W(110) on temperature appears to be
complicated.  We observe signs of a crossover at 170~K: beyond this
temperature, the rate of change of XLD spectra with temperature drops
(Sec.~\ref{sec:tempxld}, Fig.~\ref{fig:T-dependent-xmld}) and the BR
starts to vary (Sec.~\ref{sec:BR}, Fig.~\ref{fig:branching-ratio}).
On the other hand, the temperature dependence of field-induced XMCD
does not exhibit any signs of crossover (Sec.~\ref{sec:xmcd},
Fig.~\ref{fig:xmcd}).  The changes in XLD or BR at 170~K are visible
but not dramatic, and thus we do not regard them as signatures of a magnetic
phase transition.  As concerns the $T = 300$~K case, which is the
highest temperature we explored, we have strong indications that there
are local magnetic moments at Mn atoms (Sec.~\ref{sec:tempxld},
Fig.~\ref{fig:dlm}) and that these moments are not disordered as in a
paramagnetic system (Sec.~\ref{sec:tempxld}, Fig.~\ref{fig:dlm} and
Sec.~\ref{sec:xmcd}, Fig.~\ref{fig:xmcd}).

Multiscale calculations of Hasselberg \ea\ \cite{Hasselberg2015} do
not find any intermediate magnetic state between the CSS ground state
and the state without magnetic order.  In their calculations the
crossover temperature between the ordered and the disorder state was
found to be around 510~K \cite{Hasselberg2015}, which is significantly
higher than 240~K deduced from STM experiments
\cite{Sessi2009,Hasselberg2015}.  Our XAS, XLD and XMCD measurements
do not reveal a magnetic phase transition or a major change of the
magnitude of the magnetic moment per atom below 300~K, in agreement
with the predictions of Hasselberg \ea\ \cite{Hasselberg2015}.

It was suggested that the apparent loss of magnetic order seen in STM
experiments could be due to thermal depinning of the spin spirals.
Depinning effects would not affect our experimental data: XAS yields
an almost instantaneous snapshot of the magnetic state of the system,
contrary to STM, which yields information averaged over the millisecond
time scale necessary to perform typical differential conductance
measurements in a lock-in mode.  XAS thus provides information
complementary to the information obtained by STM.

Based on our data, it is difficult to assess the significance of the
crossover observed in the temperature dependence of XLD and BR around
170~K.  Possibly, an important factor can be that our samples contain
Mn monolayer stripes of different widths in the [001] direction and
that x-ray absorption spectroscopy probes many of them at the same
time.  Sessi \ea\ \cite{Sessi2009} found that magnetic properties of
Mn stripes depend on these widths.  It is conceivable that different
parts of the sample are in a different magnetic state and that this
state varies with the temperature depending on the size of the
respective Mn island.  The measured signal would then be a
superposition of different signals.  One should also have in mind that
XAS will stress the role of large Mn monolayer terraces (in comparison
to short terraces) because that is where most of Mn atoms are
sitting. Open questions clearly remain.


\section{Conclusions \label{sec:conclusions}}

X-ray linear dichroism can be used as a probe for noncollinear
magnetic order.  Cycloidal and helical spin spirals give rise to
significantly different \mnl\ XLD signals for Mn/W(110), enabling thus
to distinguish between these two configurations.  Magnetic ground
state of Mn/W(110) is an AFM cycloidal spin spiral.

Based on temperature-dependent XAS, XLD and field-induced XMCD spectra
we deduce that the magnetic order of Mn/W(110) varies with temperature,
but this variation lacks a clear indication of a phase transition in
the investigated range (8--300~K). Local magnetic moments at Mn atoms
are present for temperatures up to 300~K.

A crossover exists in the temperature dependence of XAS branching ratios
and in XLD profiles around 170~K, but it is not present in XMCD data.
The ground-state magnetic order (AFM cycloidal spin spiral) appears to
be weakened and possibly partially disrupted at these temperatures,
but the system is not paramagnetic even up to 300~K.  Tentatively, the
observed trends may result from the fact that x-ray absorption
provides an instantaneous view on the magnetic properties averaged
over a set of Mn islands with a variety of sizes and shapes: for some
Mn islands, the system might be in another --- yet undetermined ---
magnetic state.


\begin{acknowledgments}
This work was supported by the GA~\v{C}R via the project 20-18725S.
Additionally, computing resources were supported by the project
CEDAMNF CZ.02.1.01/0.0/0.0/15\_003/0000358 (Ministry of Education,
Youth and Sport) and by the project "e-Infrastruktura CZ" (e-INFRA
LM2018140) provided within the program Projects of Large Research,
Development and Innovations Infrastructures. Soft x-ray experiments
were carried out at the ESRF facilities (beamline ID08) during
projects HE3406 and HE3638. We acknowledge financial support (travel,
accommodation and subsistence) by the ESRF for both projects.
Finally, we thank the DFG for financial support via the Cluster of
Excellence ``Advanced Imaging of Matter'' (EXC 2056, project ID
390715994).
\end{acknowledgments}

\clearpage 

\appendix


\section{STM images of the substrate and the sample \label{sec:stm}}

\begin{figure}
\includegraphics[width=80mm]{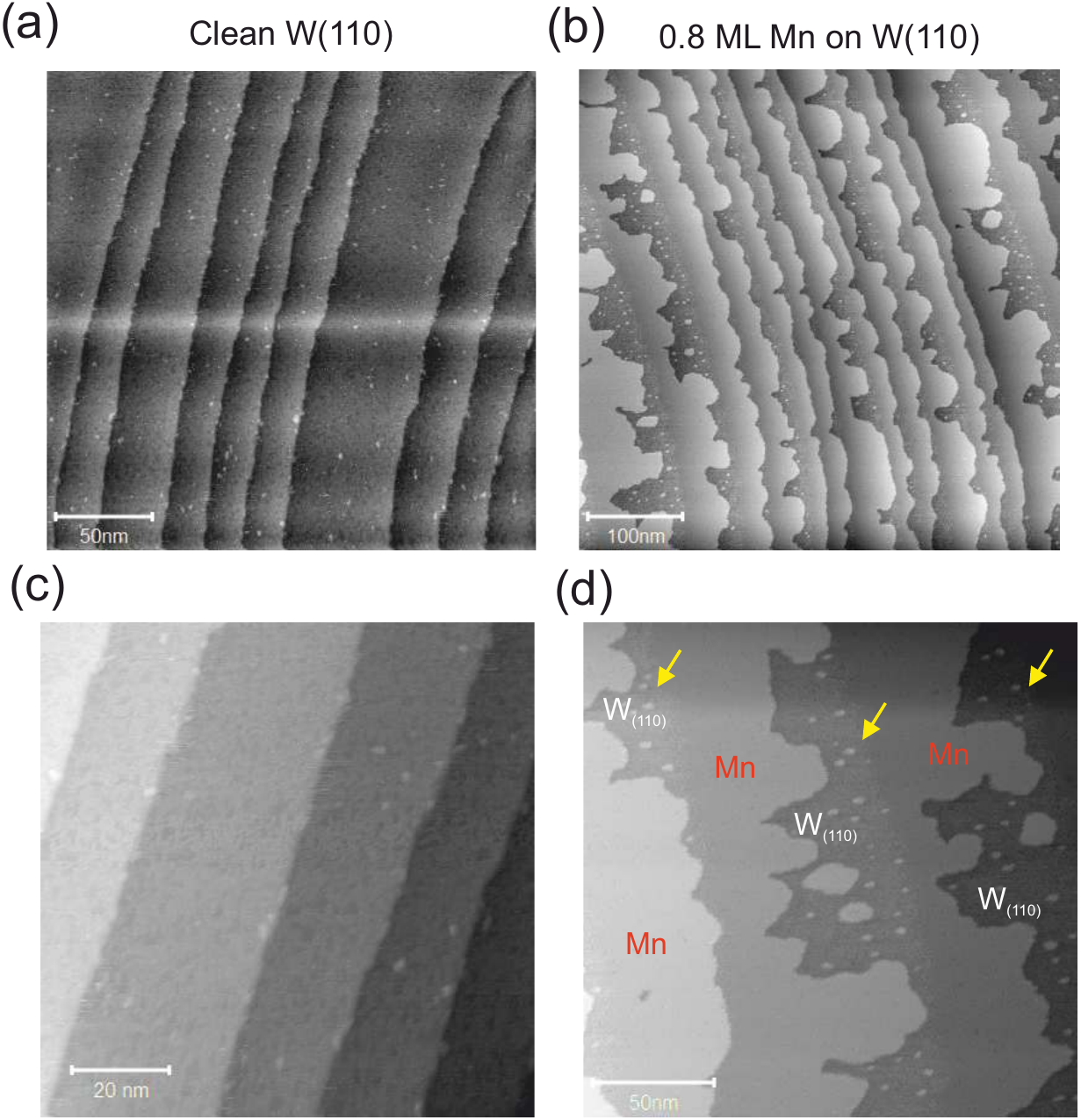}
\caption{\label{fig:STM} {\it In situ} sample characterization in the
  ID8 preparation chamber by scanning tunneling microscopy. (a)~Clean
  W(110) substrate with multiple terraces. (b)~Step-flow growth of
  monolayer Mn stripes on W(110) at Mn coverages of about 0.8 MLs. (b)
  and (d) show images of respective sample areas on a smaller
  scale. Tungsten step edges are indicated by yellow arrows.}
\end{figure}

Typical STM images of a clean W(110) surface are shown in
Fig.~\ref{fig:STM}(a) and (c). In the clean state typical
carbon-induced (15$\times$3) reconstructions have disappeared in the
LEED patterns.  STM images demonstrating the step-flow growth of Mn
monolayer stripes (referred to in Sec.~\ref{sec:sample}) are presented
in Fig.~\ref{fig:STM}(b) and (d).


\section{Internal consistency of XLD measurements \label{sec:xldcheck}}

\begin{figure}[h]
\includegraphics[width=80mm]{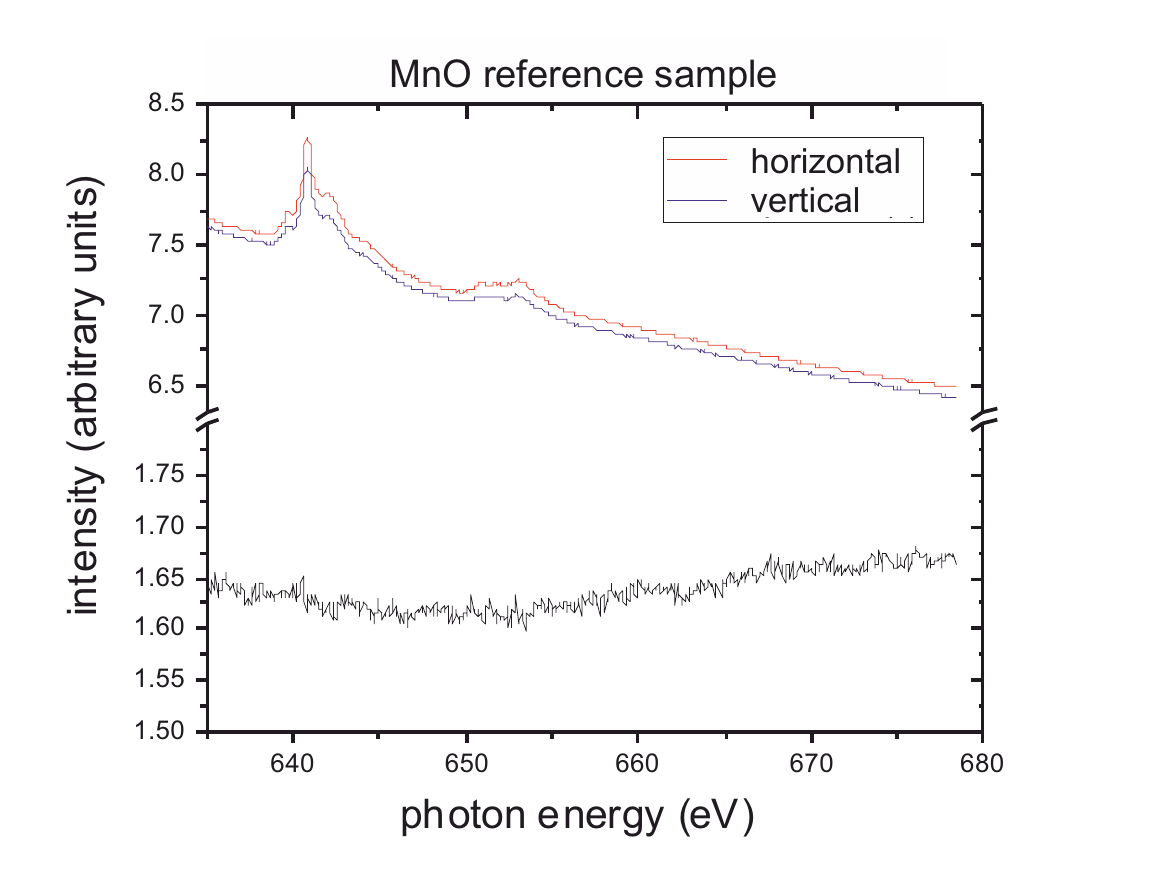}
\caption{\label{fig:MnO} XAS and XLD signals measured for the
  polycrystalline MnO reference sample. Top graph: XAS for horizontal
  and vertical photon polarization $\epsilon_{h}$ and
  $\epsilon_{v}$. Bottom graph: XLD signal resulting from the XAS
  spectra shown above. The constant offset of XLD values reflects the
  offset in the raw data in the top graph, as they are not normalized
  to 1 at 637 eV.}
\end{figure}

A typical featureless XLD signal of randomly oriented polycrystalline
MnO powder is plotted in the lower graph of Fig.~\ref{fig:MnO}
(cf.\ \cite{Gilbert2003}).  Small shifts in photon energy calibration
between measuring the spectra for $\epsilon_{h}$ and $\epsilon_{v}$
photon polarization would produce spurious nonzero XLD signals at the
Mn $L_{2,3}$ edges, with characteristic derivative character. We
comment here that indeed such artifacts appeared when the photon
polarization vector was set to angles other than the standard
$\epsilon_{h}$ and $\epsilon_{v}$ undulator configurations. Therefore,
to study the dependence of XLD on the azimuthal orientation of the
sample, we kept the polarization vectors $\epsilon_{h}$ and
$\epsilon_{v}$ constant and instead rotated the W(110) crystal.

\begin{figure}[h]
\includegraphics[width=80mm]{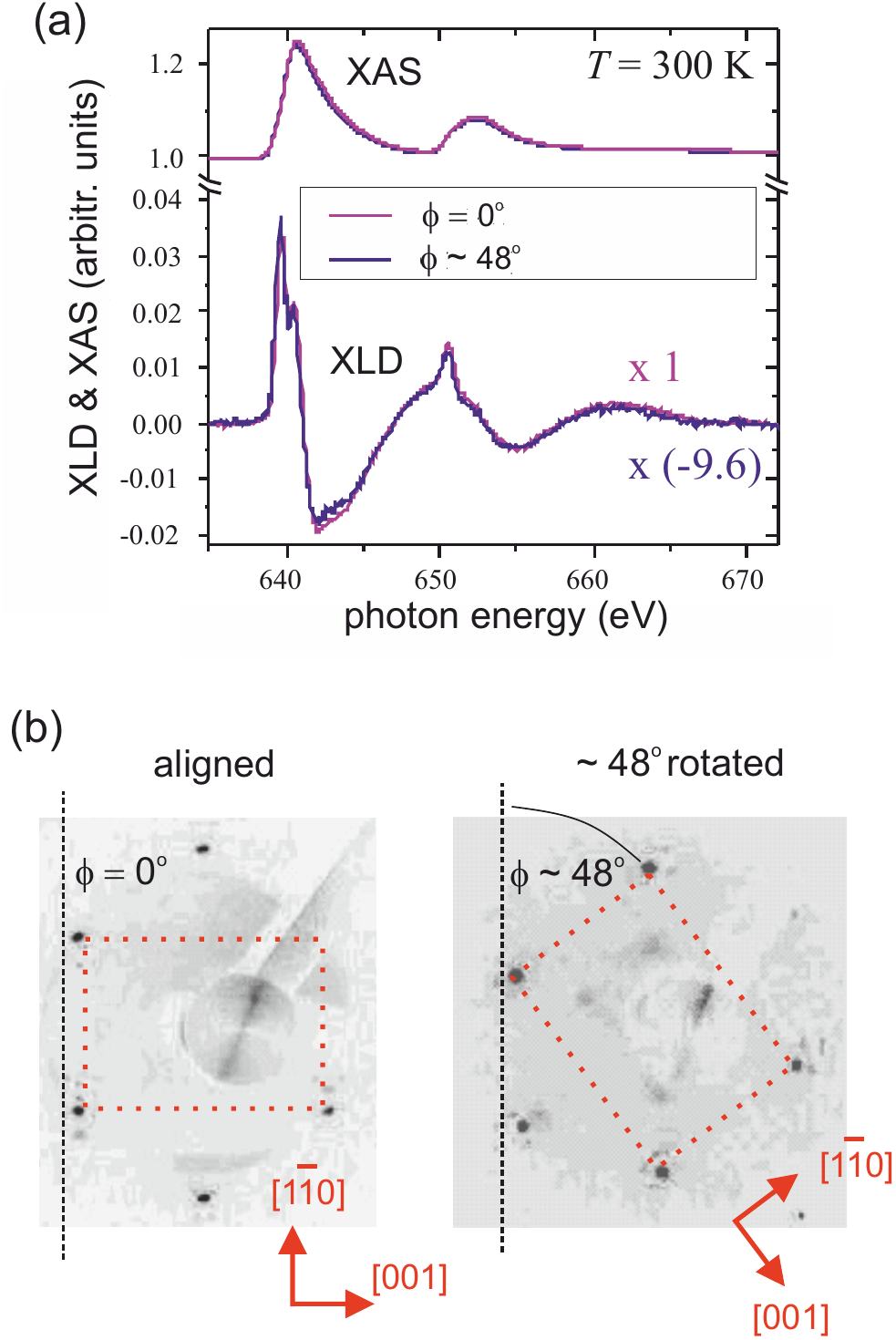}
\caption{\label{fig:aziscale} Scaling of the \mnl\ XLD intensity under
  azimuthal rotation of the W(110) substrate. All data were measured at
  $T = 300$~K. (a)~\mnl\ XAS and XLD for angles $\phi =
  0^{\circ}$ and $48^{\circ}$ between the W[1$\bar{1}$0]
  direction and the $\epsilon_{h}$ photon polarization vector.  XLD
  data for $\phi = 48^{\circ}$ are shown multiplied by
  $1/\cos(2\phi)$. (b)~LEED images used for deducing the azimuthal
  orientation of the crystal.}
\end{figure}

For a $C_{2v}$ symmetric crystal surface we expect a twofold symmetric
dependence of XAS signals versus azimuthal in-plane $\epsilon$
orientations in polar incidence geometry ($\Theta = 0^{\circ}$). The
$C_{2v}$ symmetry of the W(110) crystal surface should lead to a
fundamental dichroism between the W[100] and W[1$\bar{1}$0]
directions. For a pseudomorphically grown Mn monolayer, the spin
texture would obey this symmetry.  For a $C_{2v}$ symmetry fundamental
XLD spectra should scale as $\cos(2\phi)$, where $\phi$ is the angle
between the W[1$\bar{1}$0] high symmetry direction and the
$\epsilon_{h}$ incoming photon polarization vector \cite{LTP+11}: XLD
is largest at $\phi = 0^{\circ}$ and vanishes at $\phi =
45^{\circ}$. We verified this behavior by measuring XLD for Mn/W(110)
in two azimuthal orientations, $\phi = 0^{\circ}$ and
$48^{\circ}$. Respective data is shown in
Fig.~\ref{fig:aziscale}(a). While the average XAS intensity is
equivalent for $\phi = 0^{\circ}$ and $48^{\circ}$ (upper graph), the
XLD signal at $\phi = 48^{\circ}$ is reduced and reversed in sign
compared to that of $\phi = 0^{\circ}$.  The respective factor is
$\cos(2\times 48^{\circ})=-0.105$ [see the lower graph in
  Fig.~\ref{fig:aziscale}(a)]. The shape of the XLD, however, is the
same within the resolution of the experiment, affirming a strict
$C_{2v}$ symmetry of the system.  Respective LEED patterns in
Fig.~\ref{fig:aziscale}(b) show the azimuthal rotation of the W(110)
crystal by 48$^{\circ}$, corresponding to the $\phi = 0^{\circ}$ and
$48^{\circ}$ measuring geometries.



\section{Modeling spin spirals by averaging over AFM configurations}
\label{sec:spirals}

\begin{figure}[h]
\includegraphics[bb=10 15 258 229,width=83mm]{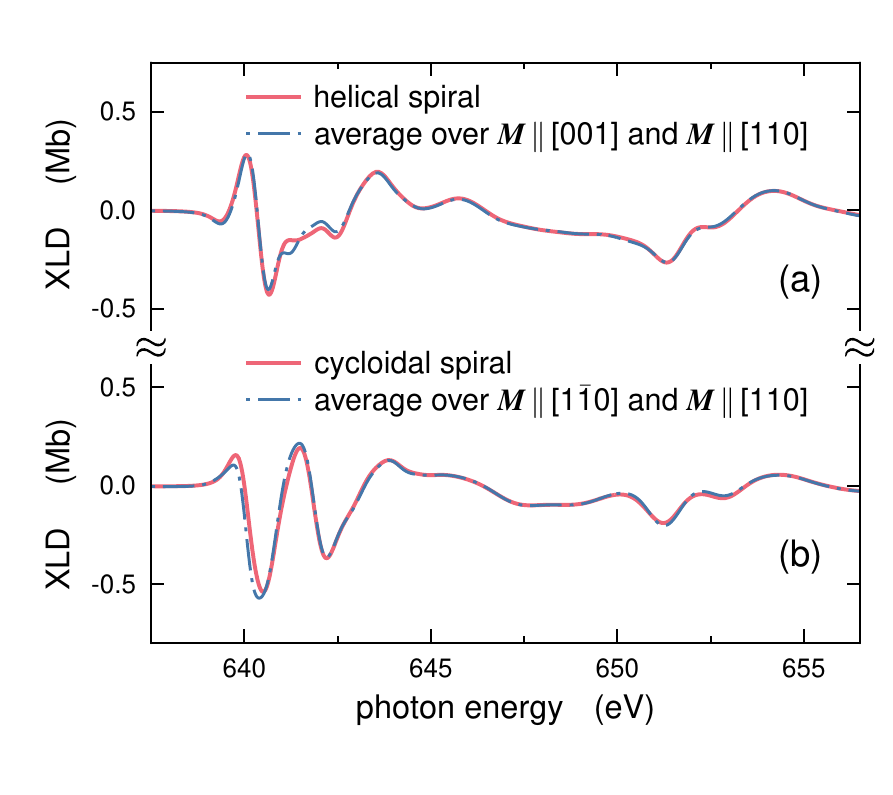}
\caption{\label{fig:spirals} Theoretical \mnl\ XLD spectra for spin
  spirals compared to averages of spectra for AFM configurations for
  two perpendicular magnetization directions.  Calculations done
  within the ASA, without a core hole. }
\end{figure}

Spectra for spin spirals are calculated as averages over spectra for
two AFM configurations, with perpendicular directions of the
magnetization.  Figure~\ref{fig:spirals} demonstrates that this
approach is justified.  We calculated the spectra for proper helical
and cycloidal spin spirals, with a wave length $\lambda = 7.2$~nm
(i.e., sixteen interatomic distances between Mn atoms along the
W[1$\bar{1}$0] direction in which the spiral propagates).  A fully
relativistic calculation is needed to account for the effect of SOC,
therefore appropriate supercells have to be involved (the generalized
Bloch theorem simplifying the calculations for spin spirals
\cite{San+91,MFE11} cannot be used if SOC is present).  This makes the
computation very demanding, so we resorted to the atomic sphere
approximation (ASA) for the potential and neglected the core hole in
this test.  The \mnl\ XLD for a helical spin spiral is compared to the
average of XLD spectra for AFM configurations with $\bm{M}\!
\parallel \! \text{W}[001]$ and with $\bm{M}\!  \parallel \!
\text{W}[110]$ in Fig.~\ref{fig:spirals}(a). The \mnl\ XLD for a
cycloidal spin spiral is compared to the average of XLD spectra for
AFM configurations with $\bm{M}\!  \parallel \!  \text{W}[1\bar{1}0]$
and with $\bm{M}\!  \parallel \!  \text{W}[110]$ in
Fig.~\ref{fig:spirals}(b).  There is nearly a perfect match between
the results obtained by both approaches.  We checked that the same
conclusions can be reached also if the average is made not just of two
but also of three or five AFM configurations with different
magnetization directions.

Our approach effectively neglects the angle between adjacent spin
directions, so it is the better the longer the spiral wave length is.
The wave length of the spin spirals observed experimentally for
Mn/W(110) is 12~nm \cite{BHB+07,Haze2017b}, which is even longer
than the wave length used in our test calculation.  Modeling spectra
of spin spirals by averages over spectra of two collinear AFM
configurations with perpendicular directions of the magnetization is
thus justified in our case.


\section{Influence of the core hole on the calculated spectra}
\label{sec:corehole}

\begin{figure}[h]
\includegraphics[bb=10 15 258 272,width=83mm]{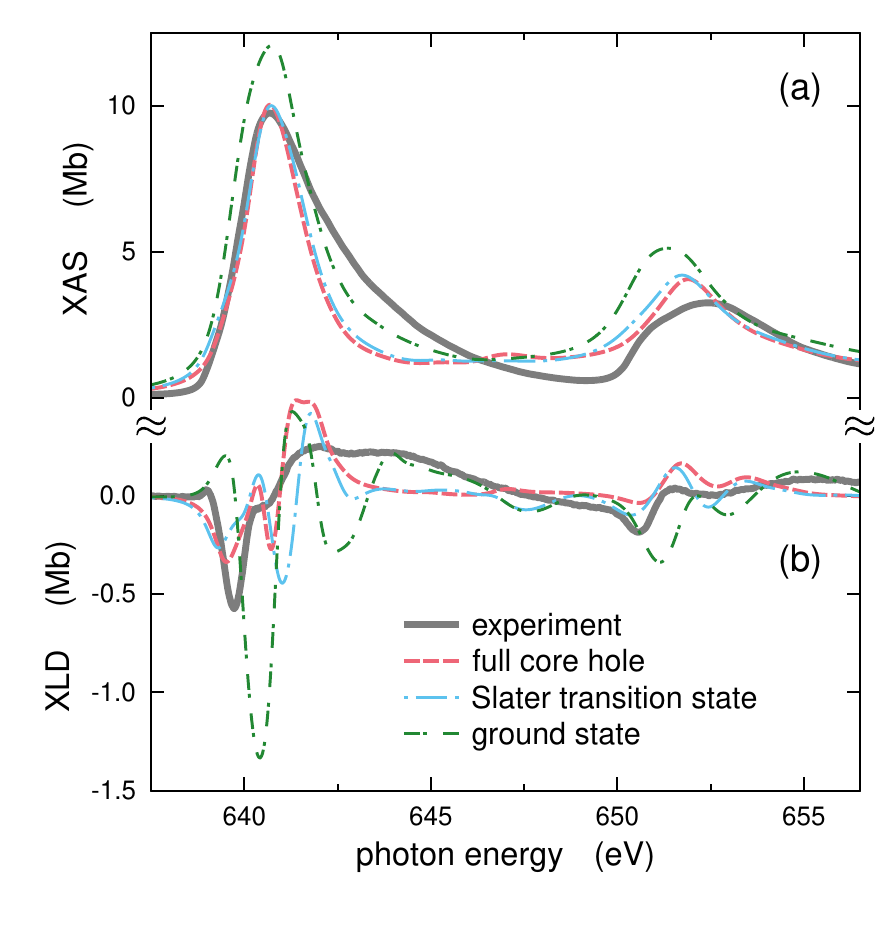}
\caption{\label{fig:core} \mnl\ XAS and XLD calculated for the ground
  state potential (core hole ignored), for the potential corresponding
  to the Slater transition state (half core hole) and for the
  potential obtained by means of the final state approximation (full
  core hole).  The experimental spectrum is shown for comparison.}
\end{figure}

In metals, the presence of the core hole is usually not the decisive
factor for the shape of XAS spectra (even though accounting for it
often improves the agreement between theory and experiment).  However,
we deal with XLD and its variations, which is quite a subtle effect 
demanding a high accuracy.  We present in Fig.~\ref{fig:core}
\mnl\ XAS and XLD spectra for a CSS magnetic configuration calculated if the
core hole is ignored (i.e., for a ground state potential), if the core
hole is included via the final state approximation, and if the Slater
transition state method is used (as the final state approximation but
with only half of the core hole).  A detailed description of the
procedures can be found in Ref.~\cite{SKJ+19}.  The presence of the
core hole does not alter the XAS spectrum substantially (similarly to
metallic Fe or Co \cite{SMS+11}), however, it makes a significant
difference for the XLD spectra.  On the other hand, there is no big
difference between the effect of a full core hole and of a half core
hole (Slater transition state).  This indicates some robustness in our
treatment of the core hole.


\section{Influence of magnetism on DOS \label{sec:dos}}

\begin{figure}
\includegraphics[bb=10 15 258 286,width=83mm]{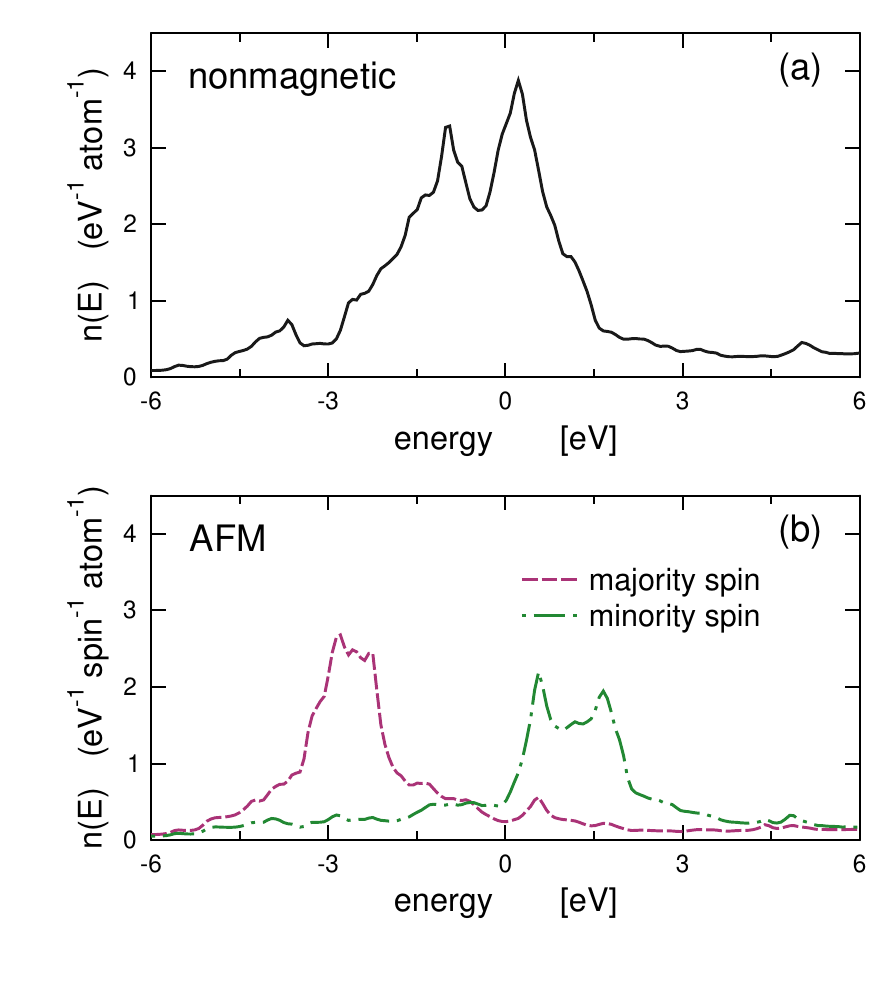} 
\caption{\label{fig:dos} DOS at Mn atoms calculated for (a) nonmagnetic
  and (b) AFM configurations of Mn/W(110).  The spin channels are shown
  separately for the AFM state and summed together for the nonmagnetic
  state. }
\end{figure}

To highlight the differences between the case with magnetic and with
nonmagnetic Mn atoms, we present in Fig.~\ref{fig:dos} the DOS at Mn
atoms for Mn/W(110) if the system is nonmagnetic and if it is in an
AFM state.  One can see that the DOS is totally different for these
two cases, similarly to what would happen if Mn was replaced by
another element.  This means, among others, that it is not possible to
separate the structural contribution to XLD from the magnetic
contribution by calculating XLD for Mn/W(110) with nonmagnetic Mn: one
would deal with a totally different system in each case.


\section{Influence of SOC on XLD \label{sec:soc}}

\begin{figure}
\includegraphics[bb=10 15 258 201,width=83mm]{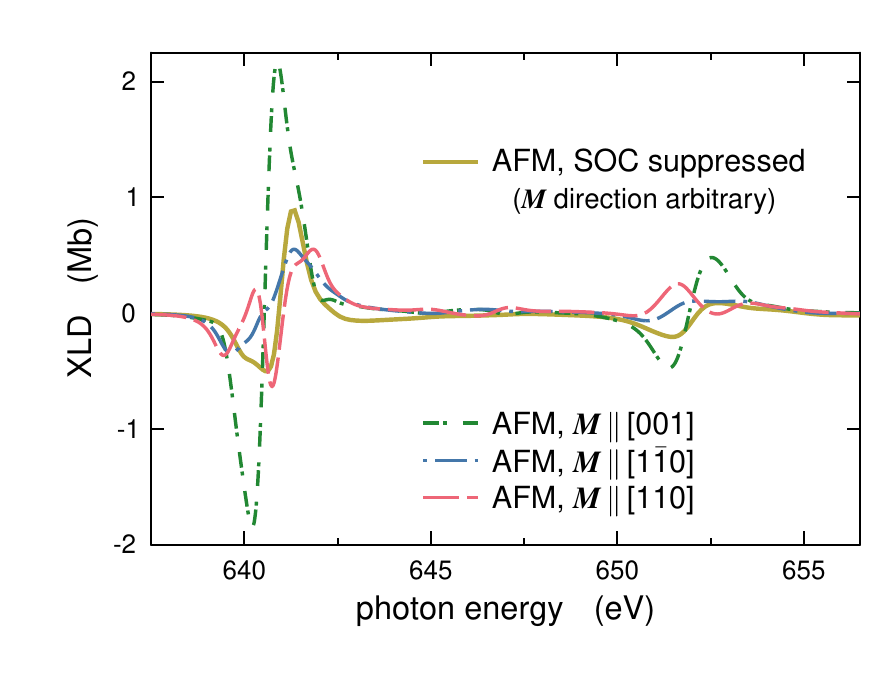} 
\caption{\label{fig:soc} \mnl\ XLD for Mn/W(110) in an AFM state, with
  the direction of magnetic moments $\bm{M}$ oriented along W[001],
  W[1$\bar{1}$0], and W[110].  An XLD spectrum calculated with SOC
  suppressed is shown for comparison. }
\end{figure}

SOC is an important factor affecting XLD of magnetic systems.
Figure~\ref{fig:soc} shows \mnl\ XLD spectra for Mn/W(110) calculated
for collinear AFM configurations, assuming that the direction of the
magnetic moments is  parallel to one of three mutually perpendicular
directions: W[001], W[1$\bar{1}$0], and W[110] (see
Fig.~\ref{fig:measuring-geometry} for a visual idea of these
directions).  One can see that the direction of the magnetic moments
(linked to the structure via SOC) has a significant impact.

It would be instructive to compare these data with results obtained
without the SOC.  This is conceptually questionable because
suppressing SOC completely (as in the scalar-relativistic formalism)
would remove the difference between the $L_{2}$ and $L_{3}$ edges 
altogether.  To suppress the SOC while keeping the distinction between
$L_{2}$ and $L_{3}$ spectra we adopt a mixed approach: (i) we suppress
the SOC for the valence states, using an approximate two-component
scheme \cite{EFVG96} employed earlier, e.g., for studying the
magnetocrystalline anisotropy \cite{SMP+16} and, additionally, (ii) we
suppress the relativistic exchange splitting of the 2$p$ core levels
pertaining to the same relativistic quantum number $\kappa$ (see,
e.g., Ref.~\cite{SVM+18} for decomposition of the $L_{3}$ and $L_{2}$
white lines into relevant components).  This second step is analogous
to the model of Kune\v{s} and Oppeneer \cite{KO03a}.  The \mnl\ XLD
spectrum calculated in this way is labeled as ``SOC suppressed'' in
Fig.~\ref{fig:soc}.  It does not depend on the direction of the
magnetic moments and can be seen as the outcome of a nonrelativistic
calculation.

It is evident that relativistic (SOC-related) effects are crucial.


\bibliography{liter-Mn-noncol}

\end{document}